\documentstyle[12pt,epsf,psfig]{article}
\begin{document}
\baselineskip 0.6cm
\newcommand{\gsim}{ \mathop{}_{\textstyle \sim}^{\textstyle >} }
\newcommand{\lsim}{ \mathop{}_{\textstyle \sim}^{\textstyle <} }
\newcommand{\vev}[1]{ \langle {#1} \rangle }
\newcommand{\EV}{ {\rm ~eV} }
\newcommand{\KEV}{ {\rm ~keV} }
\newcommand{\MEV}{ {\rm ~MeV} }
\newcommand{\GEV}{ {\rm ~GeV} }
\newcommand{\TEV}{ {\rm ~TeV} }
\newcommand{\eps}{\varepsilon}
\newcommand{\barr}[1]{ \overline{{#1}} }
\newcommand{\del}{\partial}
\newcommand{\nn}{\nonumber}
\newcommand{\ra}{\rightarrow}
\newcommand{\bino}{\tilde{\chi}}
\def\tr{\mathop{\rm tr}\nolimits}
\def\Tr{\mathop{\rm Tr}\nolimits}
\def\Re{\mathop{\rm Re}\nolimits}
\def\Im{\mathop{\rm Im}\nolimits}
\setcounter{footnote}{0}

\begin{titlepage}

\begin{flushright}
CERN-TH/2002-200\\
UT-02-48
\end{flushright}

\vskip 3cm
\begin{center}
{\large \bf  Natural Gravitino Dark Matter\\
 and Thermal Leptogenesis\\
in Gauge-Mediated Supersymmetry-Breaking Models}
\vskip 2.4cm

\center{Masaaki~Fujii$^{a,b}$ and T.~Yanagida$^{a,b,}$\footnote{
On leave from University of Tokyo.}}\\
$^{a}${\it{CERN Theory Division, CH-1211 Geneva 23, Switzerland }}\\
$^{b}${\it{Department of Physics, University of Tokyo, Tokyo 113-0033,
Japan}}\\

\vskip 2cm

\abstract{
We point out that there is no cosmological gravitino problem
in a certain class of gauge-mediated supersymmetry-breaking (GMSB) models.
The constant term in the superpotential naturally causes 
small mixings between the standard-model and messenger fields,
which give rise to late-time decays of the lightest
messenger fields. This decay provides an exquisite amount of entropy,
which dilutes the thermal relics of the gravitinos down to just
the observed mass density of the dark matter.
This remarkable phenomenon takes place naturally, 
irrespective of the gravitino mass and the reheating temperature of 
inflation, once the gravitinos and messenger fields are thermalized 
in the early Universe.
In this class of GMSB models, there is no strict upper bound on the
 reheating temperature of inflation, which makes the standard thermal
 leptogenesis  the most attractive candidate for the origin of the
 observed baryon asymmetry in the present Universe.
}

\end{center}
\end{titlepage}

\renewcommand{\thefootnote}{\arabic{footnote}}
\setcounter{footnote}{0}
%
%
%
\section{Introduction}
The minimal supersymmetric standard model (MSSM) is the most
promising candidate for physics beyond the Standard Model (SM),
since it naturally solves the ``hierarchy problem'' and leads to a
successful unification of the gauge coupling constants~\cite{coupling-unification}.
Because SUSY is not observed in the real world, it should be broken
around the TeV scale.
Once we allow generic soft SUSY-breaking terms, we must face 
hundreds of new parameters, which makes the rate of
flavour-changing neutral-current (FCNC) interactions many orders of
magnitude larger than the present experimental bounds.
In order to obtain a successful low-energy effective
theory, various mediation mechanisms of SUSY-breaking effects 
have been proposed.

The scenario most commonly considered in phenomenology
is the minimal gravity-mediated SUSY-breaking (mSUGRA) models.  This scenario
is very simple and aesthetically  attractive, since gravity does
exist in nature.
In the mSUGRA models, we also have a promising
candidate for  dark matter. That is the lightest SUSY particle (LSP),
which is usually the lightest neutralino.
The severest difficulty in this scenario is the lack of a
natural explanation for the suppression of the FCNC
interactions.
A specific form of the soft SUSY-breaking masses
must be imposed by hand in order to suppress the FCNC interactions.

It has been argued that the
suppression of the FCNC interactions can be naturally
obtained in brane-world SUSY-breaking scenarios, such as
anomaly-~\cite{AMSB} and gaugino-~\cite{gMSB} mediated SUSY-breaking models.
In these scenarios, the fields relevant to SUSY breaking are assumed to
reside on the hidden brane, which is geometrically separated from the
visible brane where the SM fields are localized.
Recently, however, a crucial observation has been made in
Ref.~\cite{Dine}, where  the authors have found
that the separation of the visible and hidden branes
in a higher-dimensional space-time
is not sufficient for suppressing the FCNC interactions.
Consequently, we need additional
{\it ad hoc} assumptions in those 
models.~\footnote{There is an attempt to realize anomaly-mediation 
models in a four-dimensional framework~\cite{s-conformal1}.}

At present, the most attractive scenario seems to be
the gauge-mediated SUSY breaking (GMSB)~\cite{GMSB}.
In the GMSB models, the suppression of the FCNC interactions is realized
in an automatic way, just because SUSY breaking occurs
at a very low-energy scale. 
Furthermore, a whole spectrum of the superparticles in the
MSSM sector is
completely determined by only a few parameters, which allows us to
discriminate the GMSB models from other candidates in the future
collider experiments.

Unfortunately, from a cosmological perspective, there exists a big difficulty
in the GMSB models. That is the so-called cosmological gravitino problem.
In these models, we have no natural explanation for the dark matter
in the present Universe.
Thermal relics of the gravitino, the LSP in the GMSB models,
overclose the Universe once they are thermalized in the early
Universe. In order to avoid the overproduction of the gravitinos,
there is a severe upper bound on the reheating temperature of inflation $T_{R}$,
which is about $T_{R}\lsim 10^6\GEV$ for $m_{3/2}=10\MEV$ for instance, and
it even reaches $T_{R}\lsim 10^3\GEV$ in the case of the lighter gravitinos
$m_{3/2}\lsim 100\KEV$~\cite{gravitino-in-GMSB}.
Furthermore, we have to fine-tune the reheating
temperature just below this upper bound to explain the required mass
density of the dark matter.  We also have to generate the observed
baryon asymmetry at just the same reheating temperature.
Therefore, for successful cosmology in the GMSB models,
we need incredible fine-tunings of various parameters,
which apparently belong to independent physics, such as SUSY breaking,
inflation and baryo/leptogenesis.~\footnote{If there exist 
extra matter multiplets
of a SUSY-invariant mass of the order of the ``$\mu$-term'',
the observed baryon asymmetry and gravitino dark matter can be 
simultaneously explained in a way totally independent of the reheating 
temperature~\cite{F-Y-extra}.}
This is a big  drawback of the GMSB models with respect to the standard
mSUGRA scenario.

In this letter, we point out that there are indeed  no such difficulties in a
certain class of GMSB models.
We consider direct gauge-mediation models where
the SUSY-breaking effects are directly transmitted to the messenger
sector without loop suppressions~\cite{direct-GMSB}.
In this class of  models, the specific form of the superpotential of the
messenger sector is usually 
provided by the $R$-symmetry: since it is violated by the constant term in the
superpotential $\vev{W}$, which is anyhow required to cancel the
cosmological constant, it is quite natural to expect that there are
small mixings
between the SM and messenger multiplets 
induced by the condensation $\vev{W}$.
As a result, the lightest messenger particle decays 
into the SM particle and gaugino through the mixings.
As we will see, the resultant late-time decays of the lightest messengers
provide an exquisite amount of
entropy, which dilutes the thermal relics of the gravitinos
down to just the observed mass density of the dark matter in the present Universe.

Surprisingly,
this miracle turns out to be true almost irrespective of the mass of the
gravitino and the reheating temperature, once the gravitinos and
messenger particles are
thermalized in the early Universe.
Consequently, by assuming 
natural mixings between the SM and messenger fields,
the severe upper bound on the reheating temperature is
completely eluded.
This fact makes it much easier to construct a realistic inflationary
scenario in the GMSB models.
This result also has an important implication on the origin of the
observed baryon asymmetry.
We will see that the standard thermal leptogenesis, through
out-of-equilibrium decays of the  right-handed Majorana neutrinos~\cite{Fuku-Yana}, 
is now the most attractive mechanism to generate the observed baryon asymmetry
in the GMSB models.

\section{Required Amount of Entropy}
In the GMSB models, the longitudinal component of the gravitino
($\sim$~Goldstino) $\psi$ interacts fairly strongly with the SM particles.
The total production cross section of $\psi$ by scattering processes is given 
by~\cite{gravitino-in-GMSB}
\begin{equation}
 \vev{\Sigma_{\rm scatt}v_{\rm rel}}\approx 5.9
\frac{g_{3}^2m_{\widetilde{G}}^2}{m_{3/2}^2M_{*}^2}\;,
\label{scattering-rate}
\end{equation}
where $\vev{~}$ denotes the thermal average; $M_{*}=2.4\times
10^{18}\GEV$ is the reduced Planck scale, $m_{\widetilde{G}}$ is the
mass of the gluino and $g_{3}$ is the coupling constant of the
SU(3)$_{\rm C}$ gauge group. Then, the  resultant interaction rate is given by
\begin{equation}
 \Gamma_{\rm scatt}\approx \vev{\Sigma_{\rm scatt}v_{\rm rel}}\;n_{\rm rad}\;,
\label{interaction-rate-scatt}
\end{equation}
with $n_{\rm rad}=\left(\zeta (3)/\pi^2\right)T^3$ being the number
density for one massless degree of freedom.

By the scattering interactions, the gravitinos  are kept in 
thermal equilibrium if
$\Gamma_{\rm scatt}/H$ $\gsim 1$,
where $H$ is the Hubble parameter of the expanding Universe. The
corresponding freeze-out temperature of $\psi$ is estimated to
\begin{equation}
 T_{f}\approx 1\TEV\left(\frac{g_{*}(T_{f})}{230}\right)^{1/2}
\left(\frac{m_{3/2}}{10\KEV}\right)^{2}\left(\frac{1\TEV}{m_{\widetilde{G}}}\right)^2\;,
\label{freeze-out-temperature}
\end{equation}
where $g_{*}(T_{f})$ denotes the effective massless degrees of freedom when
the cosmic temperature $T=T_{f}$.~\footnote{For the light gravitino $m_{3/2}\lsim
10\KEV$, the decay processes of SUSY particles are comparable
with scattering ones and keep the gravitinos in thermal equilibrium
until the temperature drops below the superparticle-mass scale.}
If the reheating temperature of inflation $T_{R}$ is higher than
$T_{f}$,  gravitinos are in thermal equilibrium in the early Universe.
Here, we have assumed the radiation-dominated Universe at the freeze-out 
time
of the gravitinos. We will justify this assumption later,  
even in the presence of the messenger particles.

If there is no additional entropy production, the resultant yield of the
thermal gravitino is estimated to
\begin{equation}
 Y_{3/2}\left(\equiv \frac{n_{3/2}}{s}\right)=\frac{45}{2 \pi^2 g_{*}(T_{f})}\frac{\zeta(3)}{\pi^2}
\left(\frac{3}{2}\right)\;,
\label{number-density}
\end{equation}
where $n_{3/2}$ is the number density of gravitinos and $s$ is the
entropy density. In terms of the density parameter, it is written as
\begin{equation}
 \Omega_{3/2}h^2\simeq 5.0 \times \left(\frac{m_{3/2}}{10\KEV}\right)
\left(\frac{230}{g_{*}(T_{f})}\right)\;,
\label{density-parameter}
\end{equation}
where $h$ is the present Hubble parameter in units of $100\;\rm{km\;
sec^{-1}\; Mpc^{-1}}$, and $\Omega_{3/2}\equiv \rho_{3/2}/\rho_{c}$;
$\rho_{3/2}$ and $\rho_{c}$ are the energy density of the gravitino and
the critical density in the present Universe, respectively.
Since the observed mass density of the dark matter is
$\Omega_{\rm DM}h^2\simeq 0.1\mbox{--}0.2$, the required dilution factor via the
late-time entropy
production is given by
\begin{equation}
 \Delta\simeq 33\times \left(\frac{m_{3/2}}{10\KEV}\right)
\left(\frac{230}{g_{*}(T_{f})}\right)\left(\frac{0.15}{\Omega_{\rm DM}h^2}\right)\;.
\label{required-entropy-production}
\end{equation}
The above entropy should be supplied after the freeze-out
time of the gravitinos.

\section{Decay of the Lightest Messenger and ${\bf\Omega_{3/2}}$}
In this section, we discuss the decay of the lightest messenger and
the resultant mass density of gravitino dark matter.
Further constraints will be discussed in the next section.
In this letter, we adopt the simplest messenger sector, which consists of
a pair of chiral supermultiplets $\Phi+\bar{\Phi}$,
describing a Dirac fermion of mass $M$ and two complex
scalar fields of mass squared $M^2\pm F$, where $F$ is the $F$-term
SUSY-breaking
component of the mass of the messenger multiplets.
In order to preserve the success of the gauge-coupling unification,
we assume that $\Phi+\bar{\Phi}$ transform as $\bf{5+\bar{5}}$ under the
SU(5)$_{\rm GUT}$ gauge group.~\footnote{Adopting a pair of
$\bf{10+\barr{10}}$ messenger multiplets does not change 
the basic results in this letter.}

Under this setup, the gauginos and SUSY particles in the MSSM sector
obtain the following soft SUSY-breaking masses,
$M_{a}$ and $m^{2}_{\rm soft}$, via gauge interactions at
the one- and two-loop level, respectively:
\begin{eqnarray}
M_{a}\simeq\frac{g_{a}^2}{16\pi^2}\Lambda\;,
\qquad
 m^2_{\rm soft}\simeq 2\sum_{a}\left\{C_{a}\left(\frac{g_{a}^2}{16\pi^2}\right)^2\right\}
\Lambda^2\;,
\label{soft-mass}
\end{eqnarray}
where $g_{a}\;(a=1,\;2\;,3)$ are gauge coupling constants and $C_{a}$
are the quadratic Casimir.
Here, $\Lambda\equiv F/M$ determines the overall scale of the soft 
SUSY-breaking masses; $\Lambda\simeq 10^{5}\GEV$ is required to obtain the
correct size of soft masses.
The mass of the messenger particles $M$ and the
gravitino mass $m_{3/2}$ are related by the 
formula:
\begin{equation}
M=\frac{\sqrt{3}\;k M_{*}}{\Lambda}m_{3/2}\simeq 4.2\times 10^{8}\;
k\left(\frac{m_{3/2}}{10\KEV}\right)\left(\frac{10^{5}\GEV}{\Lambda}\right)\GEV\;.
\label{messenger-mass}
\end{equation}
Here, we have defined $k\equiv F/F_{\rm DSB}\leq 1$, where $F_{\rm DSB}$
denotes the original
$F$-term in the dynamical SUSY-breaking sector. In direct
gauge-mediation models, this ratio is not loop-suppressed
and naturally given by $k\lsim1$.

In the present work, we assume that there is the following mixing term between
the SM and messenger multiplets through the
small $R$-symmetry-breaking effects caused by the constant term in the
superpotential, $\vev{W}$:~\footnote{Here, we assume, for instance, the
$R$-charge for $\bf \bar{5}$ and $\bf 5_{M}$ to be $+1$ and $-1$, respectively.}
\begin{equation}
 \delta W=f\;\frac{\vev{W}}{M_{*}^2}{\bf 5}_{M}{\bf \bar{5}}=fm_{3/2}{\bf 5}_{M}{\bf \bar{5}}\;,
\label{mixing-term}
\end{equation}
where $f$ denotes some unknown coefficient of the order of $1$ and the
subscript $M$ denotes the multiplet that belongs to the messenger
sector.  In the present scenario, the
lightest messenger superparticle  is most likely the scalar component of a
weak doublet.
By virtue of the above small mixing term, the lightest
messenger can decay into a SM lepton and a gaugino.~\footnote{We have
neglected the decay channels through  small Yukawa interactions.}
The decay rate is estimated to be
\begin{equation}
 \Gamma_{M}\simeq \frac{g_{2}^2}{16\pi}\left(\frac{fm_{3/2}}{M}\right)^2 M\;.
\label{messenger-decay-rate}
\end{equation}

The resultant decay temperature of the lightest messenger is given by
\begin{equation}
 T_{d}\simeq 68\MEV\times \frac{f}{\sqrt{k}}\left(\frac{10}{g_{*}(T_{d})}\right)^{1/4}
\left(\frac{m_{3/2}}{10\KEV}\right)^{1/2}\left(\frac{\Lambda}{10^{5}\GEV}\right)^{1/2}\;.
\label{messenger-lifetime}
\end{equation}
Here, we have used the relation in Eq.~(\ref{messenger-mass}).
By comparing with Eq.~(\ref{freeze-out-temperature}),
one can see that the decays of the lightest messenger always take place after
the freeze-out of the gravitinos.
Therefore, the entropy production associated with the decays of the
lightest messengers dilutes the thermal relics of the gravitinos.

Now, let us estimate the amount of entropy produced by the late-time
decays of the lightest messengers. We will also justify the assumption
of the radiation-dominated Universe
made in deriving Eq.~(\ref{freeze-out-temperature}).
Assuming the stability of the lightest messengers, their relic density has
been estimated in Ref.~\cite{Giudice-Han} as
\begin{eqnarray}
&&Y_{M}\left(\equiv \frac{n_{M}}{s}\right)\approx 3.65\times
10^{-10}\left(\frac{M}{10^6\GEV}\right)\;,
\label{messenger-abundance-yield}\\
&&\Omega_{M}h^2\approx 10^{5}\left(\frac{M}{10^6\GEV}\right)^2\;,
\label{messenger-density-parameter}
\end{eqnarray}
where $n_{M}/s$ denotes the frozen-out value of the yield of the
lightest messengers and $\Omega_{M}$ is the corresponding density
parameter. After the freeze-out of the lightest messengers, the total
energy density of the Universe is given by
\begin{equation}
 \rho=\frac{\pi^2}{30}g_{*}(T)T^4+\frac{2\pi^2}{45}g_{*}(T)T^3 M Y_{M}\;,
\label{total-energy-density}
\end{equation}
where the first and second terms represent the contributions from the
radiation and from the lightest messengers,
respectively. The thermal relics of the lightest messengers begin to
dominate the energy density at
\begin{equation}
 T_{C}=\frac{4}{3}MY_{M}\simeq 84\GEV\times k^2\left(\frac{10^5\GEV}{\Lambda}\right)^2
\left(\frac{m_{3/2}}{10\KEV}\right)^2\;,
\label{critical-temperature}
\end{equation}
where we have used the relation in Eq.~(\ref{messenger-mass}).
From Eqs.~(\ref{freeze-out-temperature}) and
(\ref{critical-temperature}), one can see that the matter-dominated
Universe starts well after the freeze-out time of the gravitinos, 
and hence the assumption made in the derivation of
Eq.~(\ref{freeze-out-temperature}) is justified.

By assuming the instantaneous decays of the lightest messengers, which
is accurate enough for the present purpose, we can
obtain the dilution factor from energy conservation as\footnote{
We are grateful to K.Hamaguchi for pointing out an error in the previous 
version.} 
\begin{eqnarray}
 \Delta_{M}\left(\equiv\frac{s_{\rm after}}{s_{\rm before}}\right)
&\simeq&\frac{4}{3}\frac{M Y_{M}}{T_{d}}\;\nonumber\\
&\simeq&
4.9\times 10^{2}\left(\frac{M}{10^8\GEV}\right)^{2}\left(\frac{10\MEV}{T_{d}}
\right) \;,
\label{messenger-dilution-factor}
\end{eqnarray}
where $s_{\rm before}\;(s_{\rm after})$ denotes the entropy density of
the Universe before (after) the decays of the lightest messengers.
By substituting Eqs.~(\ref{messenger-mass}) and (\ref{messenger-lifetime}) 
into Eq.~(\ref{messenger-dilution-factor}),
one can see that $\Delta_{M}$
has the correct order of magnitude of the required dilution factor $\Delta$ 
in Eq.~(\ref{required-entropy-production}).
From Eqs.~(\ref{density-parameter}), (\ref{messenger-mass}), (\ref{messenger-lifetime}) and
(\ref{messenger-dilution-factor}), 
we can derive the final gravitino abundance as follows:
\begin{eqnarray}
&& \Omega_{3/2}h^2=\left.\Omega_{3/2}h^2\right|_{\rm initial}\times \frac{1}{\Delta_{M}}
\nonumber\\
&&\quad\simeq
0.14\times f \left(\frac{10}{g_{*}(T_{d})}\right)^{1/4}\left(\frac{230}{g_{*}(T_{f})}\right)\left(\frac{\Lambda/k}{3\times 10^5\GEV}\right)^{5/2}
\left(\frac{2\KEV}{m_{3/2}}\right)^{1/2}\;.
\label{final-gravitino-abundance}
\end{eqnarray}

\begin{figure}[ht!]
 \centerline{\psfig{figure=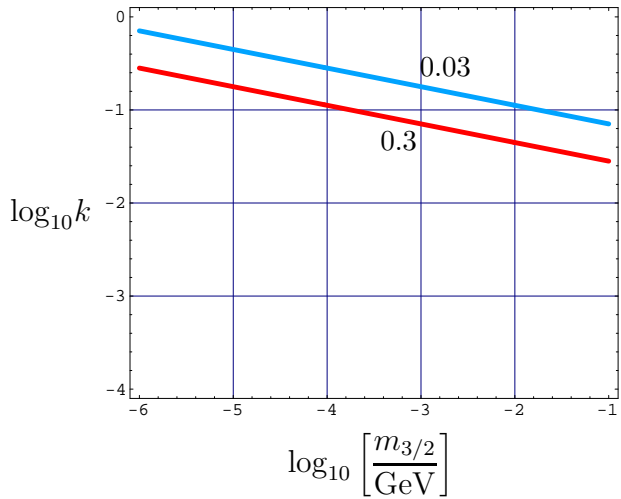,height=5.5cm}}
 \begin{picture}(0,0)
  \put(95,90){${\rm log}_{10} k$}
\put(200,-5){${\rm log}_{10}\left[\displaystyle{\frac{m_{3/2}}{{\rm GeV}}}\right]$ }
\small
\put(250,145){\bf $0.03$}
\put(235,117){\bf $0.3$}
\normalsize
 \end{picture}
\vspace{0.2cm}
 \caption{\small 
Contour plot of the abundance of the thermal gravitinos 
$\Omega_{3/2} h^2$ in a $(m_{3/2} \mbox{ -- } k)$ plane. 
The solid lines correspond to $\Omega_{3/2}h^2\simeq 0.3,\;0.03$ from 
the bottom up, respectively. Here,
we have fixed other parameters as 
$g_{*}(T_{d})=10,\;g_{*}(T_{f})=230,\;\Lambda=10^5\GEV,\;f=1$,
for simplicity.
}
\label{goodregion}
\normalsize
\end{figure}

Astonishingly, a natural parameter $k\simeq 0.1\mbox{ -- }1$ 
in direct gauge-mediation 
models and $\Lambda\approx 10^5\GEV$, which
is needed so as to obtain the correct size of the soft
SUSY-breaking masses, leads to just the mass density of the
gravitinos required to be the dominant component of 
dark matter in the present Universe.
Furthermore, the resultant abundance of gravitinos
has only a mild dependence on its mass $m_{3/2}$.
These facts can clearly be seen from Fig.~\ref{goodregion},
where we show a contour plot of $\Omega_{3/2} h^2$ in a
$(m_{3/2}\mbox{ -- }k)$ plane.
One can see that, for a given k-factor,
the gravitino masses within one order of magnitude
around a certain value 
give rise to a cosmologically interesting mass density of dark matter
$(0.03\lsim \Omega_{3/2}h^2\lsim 0.3)$.
As a result, the
gravitino dark matter is a very natural consequence of the direct
gauge-mediation models with the mixing term between the SM and messenger
multiplets given in Eq.~(\ref{mixing-term}).

Before closing this section, we briefly comment on the property of the
resultant gravitino dark matter.  The free-streaming length of the gravitino
is given by~\cite{free-streaming1}
\begin{equation}
 R_{f}\approx 0.1\left(\frac{\Omega_{3/2}h^2}{0.15}\right)^{1/3}
\left(\frac{1\KEV}{m_{3/2}}\right)^{4/3}{\rm Mpc}\;.
\label{free-streaming-scale}
\end{equation}
The late-time entropy production does not change the result in this
approximation. For a more accurate estimation, see Ref.~\cite{free-streaming2}.
The gravitino of a mass in the range $m_{3/2}\simeq
(1\mbox{--}1.5)\KEV$, whose free-streaming length is about
$R_{f}\simeq 0.1~{\rm Mpc}$,
is an interesting warm dark matter candidate,~\footnote{The possibility
of the gravitino warm dark matter with a small entropy production has been
discussed in Ref.~\cite{g-dark-entropy}, where the authors introduce the
extra matter of mass $M_{X}\simeq 10^{12}\GEV$ to obtain the required
lifetime of the lightest messenger particle. For their scenario to work,
we need an {\it ad hoc} tuning on $M_{X}$. Therefore, their model does
not solve the fine-tuning problem for the gravitino dark matter stressed
in the introduction of this letter.}
which may reconcile
the predictions of the cold dark matter with
observations~\cite{WDM,free-streaming2}. 
Heavier gravitinos with mass $m_{3/2}>(\mbox{a few})\KEV$
serve as the cold dark matter.
The gravitino with mass
$m_{3/2}\lsim 1\KEV$ is now disfavoured by
observations of Lyman-$\alpha$ forest~\cite{narayanan} and the history of
cosmological reionization~\cite{reionization}.

\section{Further Constraints}
In this section, we consider the
non-thermal gravitino production from decays of the next-to-lightest
superparticles (NLSPs). This contribution may spoil
the successful prediction in the previous section, since a huge number
of NLSPs are produced in the decays of the lightest messengers below
their freeze-out temperature.
In addition,
if the decay process of the NLSP takes place when $T\lsim 5\MEV$, 
it may also spoil the  success of the Big Bang neucleosynthesis 
(BBN)~\cite{BBN-constraints}.

Let us first consider the constraint from the BBN.
The NLSP, which is the bino or stau in the GMSB models, decays into its
superpartner and a gravitino, with the following decay width:
\begin{equation}
 \Gamma_{\chi}\simeq \frac{1}{48\pi}\frac{m_{\chi}^5}{m_{3/2}^2M_{*}^2}\;,
\label{NLSP-decay-width}
\end{equation}
where $m_{\chi}$ is the mass of the NLSP.  The corresponding 
decay temperature of the NLSP is given by
\begin{eqnarray}
 T_{\chi}&=&\left(\frac{90}{\pi^2 g_{*}(T_{\chi})}\right)^{1/4}\sqrt{\Gamma_{\chi} M_{*}}
\simeq 5\MEV\left(\frac{m_{\chi}}{100\GEV}\right)^{5/2}\left(\frac{1\MEV}{m_{3/2}}\right)\;.
\end{eqnarray}
In order not to spoil the success of the BBN, the decays
of the NLSPs should be completed before the onset of the BBN,
$T_{\chi}\gsim 5\MEV$. If we take the  natural range of the mass of the
NLSP, for instance $m_{\chi}\lsim 250\GEV$, $m_{3/2}\lsim
10\MEV$ is required for the success of the BBN.

Let us now estimate the contribution of the non-thermal gravitinos 
to the total mass density of the dark matter.
If $\Gamma_{M}>\Gamma_{\chi}$, the produced NLSPs
have enough time to annihilate before they decay into gravitinos.
This case corresponds to 
\begin{equation}
m_{3/2}\gsim 27\KEV \frac{1}{f^{2/3}}\left(\frac{m_{\chi}}{100\GEV}
\right)^{5/3}\left(\frac{3\times 10^{5}\GEV}{\Lambda/k}\right)^{1/3}\;.
\label{enough-annihilation}
\end{equation}
In this case, the resultant yield of the NLSPs before
they decay is given by the following value, to a good 
approximation~\cite{higgsino-wino}:
\begin{equation}
 \frac{n_{\chi}}{s}=\sqrt{\frac{45}{8\pi^2 g_{*}(T_{d})}}\frac{\vev{\sigma v}_{\chi}^{-1}}{M_{*}T_{d}}\;,
\label{frozen-value-of-the-NLSP}
\end{equation}
where $\vev{\sigma v}_{\chi}$ denotes the $s$-wave annihilation cross
section of the NLSP. The subsequent decays of the NLSPs produce the same
number of gravitinos. Therefore, combined with
Eq.~(\ref{messenger-lifetime}), the resultant abundance of the
non-thermal gravitinos is given by
\begin{equation}
 \Omega_{3/2}^{\rm NT}h^2\simeq 1.3\times 10^{-2}\frac{\sqrt{k}}{f}
\left(\frac{10}{g_{*}(T_{d})}\right)^{\frac{1}{4}}\left(\frac{m_{3/2}}{10\MEV}\right)^{\frac{1}{2}}
\left(\frac{10^5\GEV}{\Lambda}\right)^{\frac{1}{2}}\left(
\frac{10^{-11}\GEV^{-2}}{\vev{\sigma v}_{\chi}}\right)\;.
\label{abundance-with-annihilation}
\end{equation}
In the case of the stau NLSP, the $s$-wave annihilation cross section is
approximately given by $\vev{\sigma v}_{\chi}\simeq
10^{-7}\GEV^{-2}\left(100\GEV/m_{\chi}\right)^2$. Even in the case of
the bino NLSP, $\vev{\sigma v}_{\chi}\gsim 10^{-11}\GEV^{-2}$ is
naturally obtained for relatively large ${\rm tan}\beta$,
which is about ${\rm tan}\beta$ $\gsim 15\;(30)$ for $m_{\chi}\simeq 
100\; (250)\GEV$, 
via small higgsino contamination. 

\begin{figure}[ht!]
 \centerline{\psfig{figure=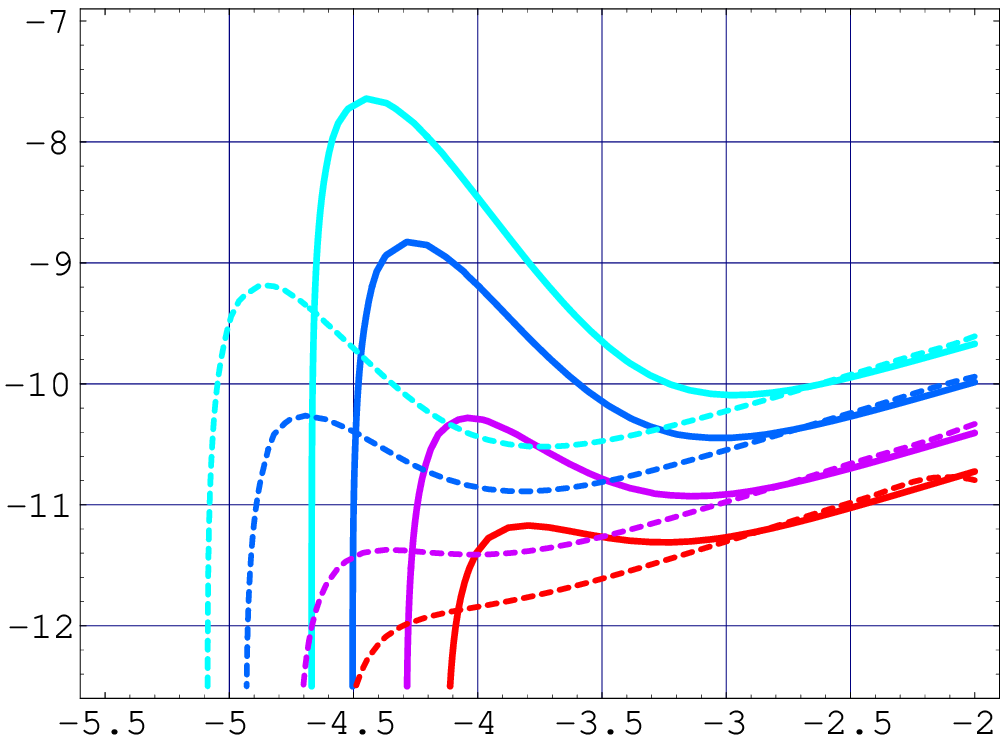,height=5.5cm}}
 \begin{picture}(0,0)
  \put(45,100){${\rm log}_{10}\left[
\displaystyle{\frac{\vev{\sigma v}_{\chi}}{{\rm GeV}^{-2}}}\right]$}
\put(200,-5){${\rm log}_{10}\left[\displaystyle{\frac{m_{3/2}}{{\rm GeV}}}\right]$ }
\small
\put(340,100){\bf $1\%$}
\put(340,90){\bf $2\%$}
\put(340,80){\bf $5\%$}
\put(335,70){\bf $10\%$}
\normalsize
 \end{picture}
\vspace{0.2cm}
 \caption{\small Contour plot of the abundance of the non-thermal gravitinos 
$\Omega_{3/2}^{\rm NT}h^2$. Solid (dashed) 
lines correspond to $\Omega_{3/2}^{\rm NT}h^2=10$, $5$, $2$ and $1\%$ 
of the total mass density of the dark matter (
$\Omega_{\rm DM}h^2$ $\simeq 0.15$), for $m_{\chi}=250\;(100)\GEV$, 
from the bottom up, respectively. The previous estimation given 
in Eq.~(\ref{abundance-with-annihilation}) is valid in the region where 
the contour lines are almost straight.
In this calculation, we have fixed $k=1$. 
}
\label{abundance}
\normalsize
\end{figure}

The above estimation is not valid if $\Gamma_{M}<\Gamma_{\chi}$,
since the produced NLSPs immediately decay into  gravitinos. 
In this case, we have to follow the full evolution by solving
coupled Boltzmann equations.~\footnote{
Significant annihilations take place also in this case.} 
The result is given in Fig.~\ref{abundance}.
In this calculation, we have conservatively assumed that the number
of NLSPs produced per decay of the lightest messenger particle
is given by $N_{\chi}\approx M/E_{\rm th}$, where $E_{\rm th}\approx m_{\chi}^2/T_{d}$ 
is the threshold energy 
to produce the NLSPs by scatterings with the thermal backgrounds 
with $T=T_{d}$.  One can see that the contribution of the 
non-thermal gravitinos is 
smaller than several per cent  of the total mass density of the dark 
matter as long as $\vev{\sigma v}_{\chi}\gsim 10^{-11}\GEV^{-2}$.

In summary, the prediction of the gravitino dark matter given in 
Eq.~(\ref{final-gravitino-abundance}) is not spoiled in most of the 
parameter region. The relevant constraint comes only from the 
BBN, which is satisfied as long as $m_{3/2}\lsim 10{\MEV}$
for a reasonable range of $m_{\chi}$.
Finally, we briefly comment on another interesting aspect. 
The non-thermal gravitinos 
produced from the NLSPs behave as a hot/warm component of 
dark matter, which contributes several per cent to the 
total mass density of dark matter in the case of the bino NLSP with 
moderate ${\rm tan}\beta$. Therefore, the present model can naturally
realize a mixed dark matter scenario.
By virtue of the late-time production of the NLSPs,
an unnaturally  large hierarchy between the slepton
and the bino masses is not required to realize
a mixed dark matter scenario~\cite{Mixed-DM}. 

\section{Thermal Leptogenesis}
As we have mentioned in the introduction, the present model
has important implications on the origin of
the baryon asymmetry in the present Universe.
By virtue of the entropy production by the decays of the lightest messengers,
we obtain the required mass density of the dark matter
without any adjustments of the reheating temperature. 
Consequently, the standard thermal 
leptogenesis~\cite{Fuku-Yana,Buch}  now becomes
the most promising candidate for the origin of the observed baryon
asymmetry in the GMSB models.

Let us start by introducing the relevant terms in the
superpotential:
\begin{equation}
 W=\frac{1}{2}M_{Ri}N_{i}N_{i}+h_{i\alpha}N_{i}L_{\alpha}H_{u}\;,
\label{superpotential}
\end{equation}
where $N_{i}\;(i=1,2,3)$ denote the heavy right-handed
Majorana neutrinos of mass $M_{Ri}$; $L_{\alpha}\;
(\alpha=e,\mu,\tau)$ and $H_{u}$ denote the lepton doublets and the
Higgs doublet that couples to up-type quarks, respectively. The
lepton-number asymmetry per decay of a right-handed neutrino $N_{i}$ is
given by~\cite{Fuku-Yana,epsiron-new}
\begin{eqnarray}
 \epsilon_{i}&\equiv&\frac{\sum_{\alpha}\Gamma(N_{i}\rightarrow L_{\alpha}+H_{u})-\sum_{\alpha}
\Gamma(N_{i}\rightarrow \barr{L_{\alpha}}+\barr{H_{u}})}{\sum_{\alpha}\Gamma(N_{i}\rightarrow
L_{\alpha}+H_{u})+\sum_{\alpha}\Gamma(N_{i}\rightarrow \barr{L_{\alpha}}+\barr{H_{u}})}\nonumber\\
\nonumber\\
&=&-\frac{1}{8\pi}\frac{1}{(hh^{\dag})_{ii}}\sum_{k\neq i}{\rm Im}\left[
\left\{(hh^{\dag})_{ik}\right\}^2\right]
\left[
{\cal F}_{V}\left(\frac{M_{Rk}^2}{M_{Ri}^2}\right)
+{\cal F}_{S}\left(\frac{M_{Rk}^2}{M_{Ri}^2}\right)
\right]\;,
\label{assymetry-parameter}
\end{eqnarray}
where $N_{i}$, $L_{\alpha}$ and $H_{u}$
($\barr{L_{\alpha}}$ and $\barr{H_{u}}$) symbolically denote fermionic
or scalar components of corresponding supermultiplets (and their
antiparticles); ${\cal F}_{V}(x)$ and ${\cal F}_{S}(x)$ represent the
contributions from vertex and self-energy diagrams, 
respectively~\cite{self-energy-vertex}:
\begin{equation}
 {\cal F}_{V}(x)=\sqrt{x}\;{\rm ln}\left(1+\frac{1}{x}\right),\qquad
{\cal F}_{S}(x)=\frac{2\sqrt{x}}{x-1}\;.
\end{equation}

For hierarchical right-handed neutrinos $M_{R1}\ll M_{R2}\;,M_{R3}$, the
lepton asymmetry is dominantly supplied by decays of the lightest
right-handed (s)neutrinos. In the following discussion, we assume that this
is the case for simplicity. In this case, the expression of the
asymmetry parameter $\epsilon_{1}$ is given by 
\begin{equation}
 \epsilon_{1}=\frac{3}{8\pi}\frac{M_{R1}m_{\nu 3}}{\vev{H_{u}^{0}}^2}\delta_{\rm eff}\;,
\label{largest-asymmetry-parameter}
\end{equation}
where $m_{\nu 3}$ is the mass of the heaviest left-handed neutrino and
$\delta_{\rm eff}$ is an effective CP-violating phase.
If the lightest right-handed (s)neutrinos are in thermal equilibrium,
the resultant lepton asymmetry is given by the following formula:
\begin{equation}
 \left|\frac{n_{L}}{s}\right|=\frac{1}{\Delta}\times\frac{45}{2\pi^2 g_{*}(T_{B})}\frac{\zeta(3)}{\pi^2}\left(\frac{3}{2}+2\right)|\epsilon_{1}|\kappa\;,
\label{lepton-asymmetry}
\end{equation}
where $T_{B}$ is the freeze-out temperature of $N_{1}$, and
$g_{*}(T_{B})\approx 270$;~\footnote{In this expression, 
we have assumed that a pair
of messenger multiplets $({\bf 5+\bar{5}})$ are in thermal equilibrium.
After the decoupling of the messenger particles, the contents of the MSSM
give rise to $g_{*}\approx 230$. If we take into account the effective
degrees of freedom in the dynamical SUSY-breaking sector, $g_{*}(T_{B})$
would be larger than this value, but it changes the resultant asymmetry
by only a factor of at most ${\cal{O}}(1)$.}
$\kappa$ denotes the fraction of the produced asymmetry that survives
washout processes by lepton-number-violating interactions after $N_{1}$
decay. Here, we have assumed
that there is no entropy production before the decays of the lightest
messengers, and set $\Delta_{M}=\Delta$ for simplicity.
For $\kappa\sim 1$,  the following out-of-equilibrium condition should be satisfied~\cite{Buch}:
\begin{equation}
 \widetilde{m}_{1}=\frac{8\pi\vev{H_{u}^{0}}^2}{M_{R1}^2}\Gamma_{N_{1}}
=(hh^{\dag})_{11}\frac{\vev{H_{u}^{0}}^2}{M_{R1}}\lsim 5\times 10^{-3}\EV\;.
\end{equation}
The lepton asymmetry produced by the right-handed (s)neutrino decays
is subsequently converted into the baryon
asymmetry by the sphaleron effects:
\begin{equation}
 \frac{n_{B}}{s}=C\frac{n_{L}}{s}\;,
\label{baryon-asymmetry}
\end{equation}
where $C$ is a number of ${\cal O}(1)$, which takes the value
$C=-8/23$ in the MSSM.

\begin{figure}[ht!]
 \centerline{\psfig{figure=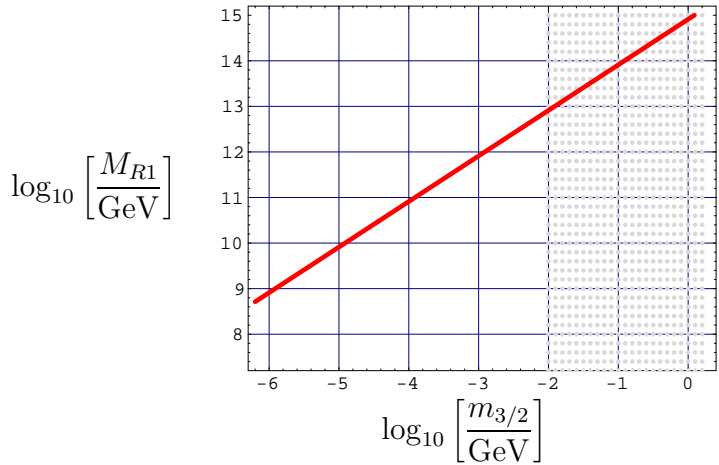,height=5.2cm}}
 \begin{picture}(0,0)
  \put(55,90){${\rm log}_{10}\left[\displaystyle{\frac{M_{R1}}{{\rm GeV}}}\right]$}
\put(195,-2){${\rm log}_{10}\left[\displaystyle{\frac{m_{3/2}}{{\rm GeV}}}\right]$ }
 \end{picture}
\vspace{0.1cm}
 \caption{
\small 
The lower bound on $M_{R1}$ (and hence $T_{R}$) to obtain the observed baryon
asymmetry. Here, we have assumed the hierarchical right-handed
neutrinos $(M_{R1}\ll M_{R2},\;M_{R3})$ and the relation given in
Eq.~(\ref{simplification-assumption}). The shaded region is disfavoured by
the BBN constraint.}
\label{lowest-MR}
\normalsize
\end{figure}

From Eqs.~(\ref{required-entropy-production}),
(\ref{largest-asymmetry-parameter}), (\ref{lepton-asymmetry}) and
(\ref{baryon-asymmetry}), we can derive the baryon asymmetry
in the present Universe.
In terms of the density parameter it is written as
\begin{equation}
 \Omega_{B}h^2\simeq 0.02\left(\frac{10\KEV}{m_{3/2}}\right)
\left(\frac{g_{*}(T_{f})}{230}\right)\left(\frac{270}{g_{*}(T_{B})}\right)
\left(\frac{M_{R1}}{10^{10}\GEV}\right)\left(\frac{m_{\nu 3}}{0.06\EV}\right)
\kappa\delta_{\rm eff}\;.
\label{baryon-density-parameter}
\end{equation}
In Fig.~\ref{lowest-MR}, we show the lower bound of $M_{R1}$ (and
hence, it is the lower bound on the reheating temperature $T_{R}$) to obtain
the required baryon asymmetry. Here, we have assumed the following
relation for simplicity:
\begin{equation}
 \left(\frac{g_{*}(T_{f})}{230}\right)\left(\frac{270}{g_{*}(T_{B})}\right)
\left(\frac{m_{\nu 3}}{0.06\EV}\right)\kappa\delta_{\rm eff}\leq 1.
\label{simplification-assumption}
\end{equation}
As can be seen, the observed baryon asymmetry $\Omega_{B}h^2\simeq 0.02$
can be naturally explained by the thermal leptogenesis in a wide range
of the gravitino mass.~\footnote{In the simplest chaotic inflationary
scenario, for example, ${\cal O}(1)$ couplings of the inflaton to the SM
fields lead to $T_{R}\approx 10^{13}\GEV$.}
From Eq.~(\ref{messenger-mass}), one can see that the required lower
mass bound on the right-handed neutrino is always larger than the mass
of the messenger fields, and then the thermalization condition for the
messenger particles is always satisfied. 
We should stress
that we do not  have to fine-tune the couplings of the inflaton to the SM
particles to reduce the reheating temperature, which is usually imposed
in SUSY inflationary models because of the cosmological gravitino problem.
We should also stress that
there is an interesting possibility to determine 
$m_{3/2}$ directly in future collider experiments
if the gravitino is lighter than about $100\KEV$~\cite{maki-orito}.
Note that such a light gravitino has been
considered as unlikely because of the cosmological gravitino problem.
However, as we have seen, it is now indeed {\it
favoured} from cosmological perspectives.
\section{Conclusions and Discussion}
In this letter, we have pointed out that there is no fine-tuning problem
to obtain the required mass density of the dark matter
in a certain class of GMSB models. By virtue of the small mixing
between the SM fields, which is induced by the 
$R$-symmetry-breaking effects, the lightest messenger particle has a 
finite lifetime and provides an exquisite amount of entropy, which
dilutes the thermal relics of the gravitinos down to just the 
mass density required for  the dark matter. This phenomenon takes place 
naturally, regardless
of the gravitino mass and the reheating temperature of inflation
as long as the gravitinos and messenger fields are thermalized in the early
Universe. There is no severe upper bound on the reheating temperature in
this class of GMSB models, which makes the standard thermal leptogenesis
very attractive as the origin of the observed baryon asymmetry.

The present scenario should have important implications also on other 
candidates for the origin of the present baryon asymmetry.
It would be very interesting to reanalyse
those models with the disappearance of the cosmological
gravitino problem taken into account.

\small
\section*{Acknowledgements}
The authors wish to thank K.Hamaguchi for 
a useful discussion.
M.F. thanks the Japan Society for the Promotion of 
Science for financial support.
This work was partially supported by Grant-in-Aid for Scientific Research
(S) $14102004$ (T.Y.).


\end{document}